\documentclass[pdflatex,sn-mathphys-num]{sn-jnl}

\usepackage{graphicx}%
\usepackage{multirow}%
\usepackage{amsmath,amssymb,amsfonts}%
\usepackage{amsthm}%
\usepackage{mathrsfs}%
\usepackage[title]{appendix}%
\usepackage{xcolor}%
\usepackage{textcomp}%
\usepackage{manyfoot}%
\usepackage{booktabs}%
\usepackage{algorithm}%
\usepackage{algorithmicx}%
\usepackage{algpseudocode}%
\usepackage{listings}%
\usepackage{lineno} 

\theoremstyle{thmstyleone}%
\theoremstyle{thmstyletwo}%

\theoremstyle{thmstylethree}%

\begin{document}

\title[Article Title]{Correlation Swapping: a correlator for independent thermal light sources}







\author[1]{\fnm{Wanting} \sur{Hou}}
\author[1]{\fnm{Jilun} \sur{Zhao}}

\author*[2]{\fnm{Zhiyuan} \sur{Ye}}\email{yezy@bnu.edu.cn}

\author[3]{\fnm{Hong-Chao} \sur{Liu}}

\author[1]{\fnm{Hai-Bo} \sur{Wang}}

\author*[1]{\fnm{Jun} \sur{Xiong}}\email{junxiong@bnu.edu.cn}

\author*[1]{\fnm{Kaige} \sur{Wang}}\email{wangkg@bnu.edu.cn}

\affil[1]{\orgdiv{School of Physics and Astronomy, Applied Optics Beijing Area Major Laboratory}, \orgname{Beijing Normal University}, \orgaddress{\city{Beijing}, \postcode{100875}, \country{China}}}

\affil[2]{\orgdiv{Department of Physics, Faculty of Arts and Sciences}, \orgname{Beijing Normal University, Zhuhai}, \orgaddress{\city{Zhuhai}, \postcode{519087}, \country{China}}}

\affil[3]{\orgname{Macao Centre for Research and Development in Advanced Materials, Institute of Applied Physics and Materials Engineering, University of Macau}, \city{Taipa, Macao S.A.R.}, \postcode{999078}, \country{China}}

\abstract{Optical intensity correlation is a fundamental property of light and an essential resource for numerous optical applications. In this work, we introduce the concept of classical correlation swapping, a classical analogue of quantum entanglement swapping, to generate all-purpose spatial correlations between two independent thermal light sources. Using a spatially unresolved Mach-Zehnder interferometer and a medium variable, we theoretically and experimentally demonstrate the feasibility of the correlator with two distinct schemes for independent pseudo-thermal light beams. Notably, the resulting photon correlations can be readily tailored to exhibit either bunching (peak) or anti-correlated (dip) characteristics, with a significantly reduced number of post-selection measurements. Leveraging this classical correlator, we further demonstrate the first classical ghost imaging experiment using uncorrelated or unknown light. Numerical simulations also confirm that the correlator can not only operate well in other spatial degrees of freedom (e.g., orbital angular momentum) but also be used to establish specific spatial correlations between pseudo-thermal light sources possessing distinct correlation properties. This work opens a new avenue for harnessing uncorrelated classical light in correlation-based optical applications.
}




\maketitle

\section{Introduction}\label{sec1}
Classical correlation of light is a crucial and fundamental phenomenon in physics. The optical intensity interferometer was first proposed in 1956 by Hanbury Brown and Twiss (HBT) for measuring stellar angular diameter \cite{HBT1956,Brown1956}. The core idea is that for a thermal light source, the intensity fluctuations measured at two spatially separated detectors are correlated within a certain space. 
The degree of this correlation, quantified by the second-order correlation function $g^{(2)}$, decreases as the distance between the two detectors increases. By measuring how this correlation changes with detector separation, one can determine the angular size of the source. The HBT interferometer is a fundamental tool for characterizing light sources, distinguishing between thermal light ($g^{(2)}(0)>1$), coherent laser light ($g^{(2)}(0)=1$), and non-classical light ($g^{(2)}(0)<1$). This technique also finds broad applications in astronomy \cite{Abeysekara2020} and other branches of physics, such as particle physics \cite{Wiedemann1999}.

At the beginning of this century, the intensity correlation of thermal light was applied to ghost imaging (GI), ghost interference, and subwavelength interference, etc., similar to a quantum entangled source \cite{Bennink2002,Bennink2004,Gatti2004,Brambilla2004,Cheng2004,Wang2004,Cai2004,Ferri2005,Cao2005,Valencia2005,Xiong2005,Zhang2005,ChenLX2021,XiaoT2023,YEZY2025,LiP2024,Forbes2025}. In these typical applications, the thermal light beam is usually divided into two correlated parts by a beamsplitter (BS). 
As for two independent thermal sources, while it is possible to observe two-photon interference \cite{Paul1986,ZhaiYHc2006,Oppelc2012,Ihn2017,Wang2023} in their superposed field, it is fundamentally impossible to observe any intensity correlation between them when their intensities are measured individually. This would limit the application of classical correlation effects in a complicated optical system that includes multiple uncorrelated or decorrelated thermal sources. 

In the 1990s, numerous experiments demonstrated that two pairs of entangled photons produced from independent quantum sources can establish quantum correlation through a technique known as entanglement swapping \cite{Zukowski1993,Pan1998}. Entanglement swapping is based on a Bell-state measurement of two idle photons, which projects the two signal photons onto an entangled state. In 2019, Bornman \textit{et al.} \cite{Bornman2019} performed the first GI experiment using entanglement-swapped photons, albeit with only a four-pixel object resolved. Recently, Qiu \textit{et al.} \cite{Qiu2023} realized the remote transport of high-dimensional orbital angular momentum (OAM) states \cite{MairA2001,LeachJ2010,Maga2016} and further demonstrated GI experiments with interaction-free light \cite{ZHangY2019,YangY2023} in the OAM space. In analogy with quantum entanglement swapping, is it possible to establish intensity correlation between two independent classical light sources?

Inspired by quantum entanglement swapping, in this paper, we propose a classical correlator capable of generating spatial intensity correlation between two uncorrelated thermal light beams. Two observation schemes for this correlator are designed within a \textit{spatially-unresolved} Mach-Zehnder interferometer. Both theoretical and experimental results confirm its operational principle, despite the currently weak correlation strength. In this context, we further demonstrate the first classical GI using uncorrelated and interaction-free light.
This correlator can be regarded as a classical analogue of quantum entanglement swapping and is expected to find specific applications in correlation optics, allowing a remote generation and/or manipulation of transverse correlations between two independent thermal light sources even possessing distinct correlation properties.



\section{Results}\label{sec2}
\subsection{Overview of the optical layout}
 \label{sec:2:1}

\begin{figure}[htbp]
\centering
{\includegraphics[width=0.9\linewidth]{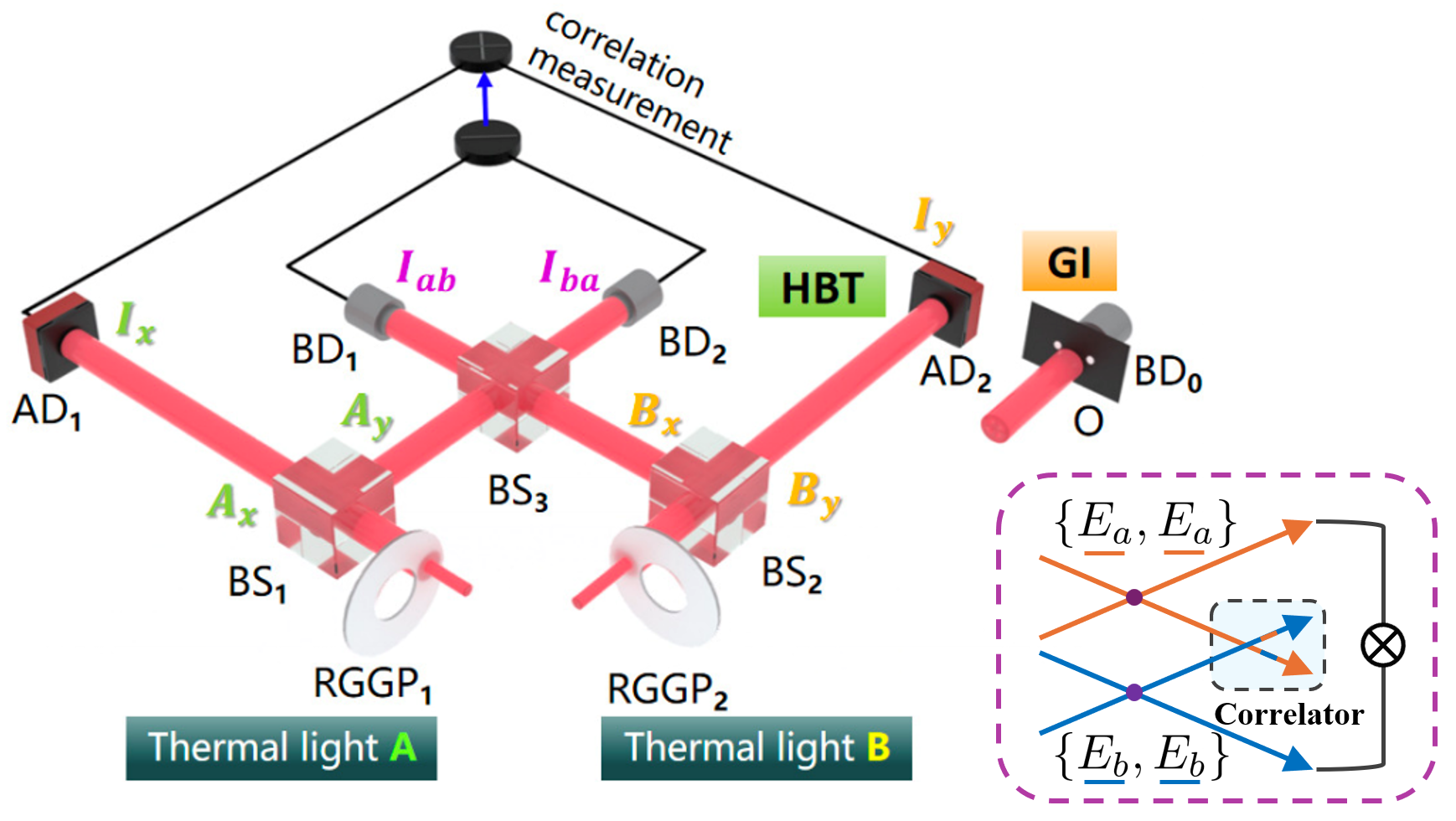}}
\caption{Schematic diagram of a correlator for classical light. BS: 50/50 non-polarizing beamsplitter, BD: bucket detector, AD: array detector, O: object, RGGP: rotating ground glass plate. The dashed box in the lower right corner shows a simplified schematic diagram of this classical correlator.}
\label{fig1}
\end{figure}

Referring to the quantum model, our basic strategy is to take a portion of two uncorrelated beams to set up a \textit{medium} measurement. The schematic diagram of the correlator is shown in Fig.~\ref{fig1}. Although true thermal light is very common in nature, its extremely short coherence time \cite{Zhang2005,Shevchenko2015,Shevchenko2017} often approaches or even exceeds the bandwidth limit of the joint detection unit. Therefore, in most practical applications related to correlation optics, pseudo-thermal light sources with a longer equivalent coherence time are predominantly used. The correlator proposed in this paper uses pseudo-thermal light sources to perform proof-of-principle experiments, but in principle it is applicable to true thermal light as long as the detection bandwidth is sufficiently large.

\begin{figure}[htbp]
\centering
{\includegraphics[width=\linewidth]{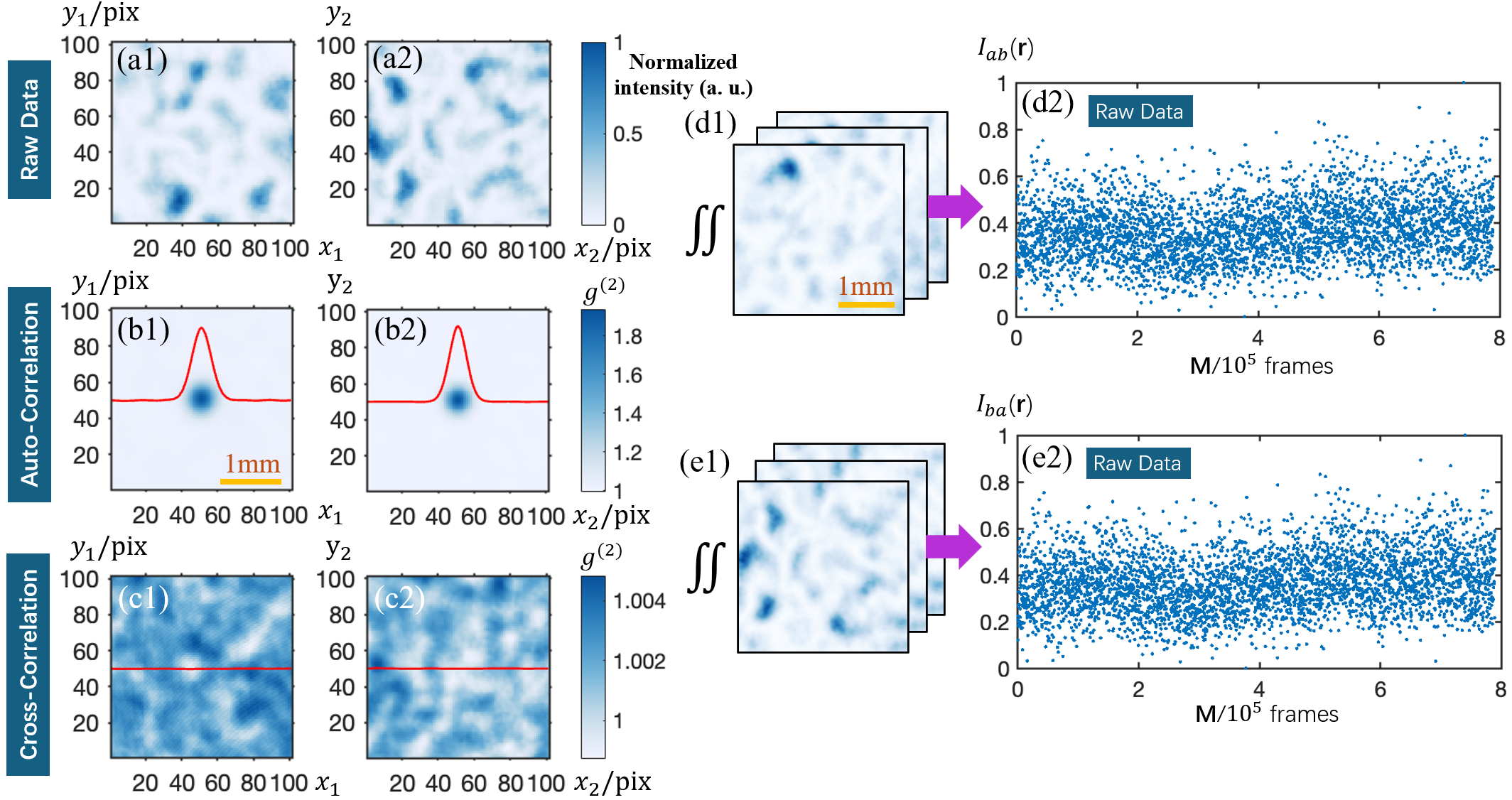}}
\caption{Experimental observations of spatial correlations between two independent pseudo-thermal light sources. (a1)(a2) Instantaneous intensity profiles of the two independent pseudo-thermal sources, $I_x(\mathbf{r}_x)$ and $I_y(\mathbf{r}_y)$; (b1)(b2) Auto-correlation measurement, $g^{(2)}[I_x(\mathbf{r}_0),I_x(\mathbf{r})]$ and $g^{(2)}[I_y(\mathbf{r}_0),I_y(\mathbf{r}_y)]$; (c1)(c2) Cross-correlation measurement, $g^{(2)}[I_y(\mathbf{r}_0),I_x(\mathbf{r}_x)]$ and $g^{(2)}[I_x(\mathbf{r}_0),I_y(\mathbf{r}_y)]$; The red solid line superimposed on the figure represents the one-dimensional profile of the central region. (d)(e) show the raw experimental data obtained from the correlator. (d1)(e1) Captured instantaneous speckle patterns, $I_{ab}(\mathbf{r})$ and $I_{ba}(\mathbf{r})$, at the outports of BS$_3$. (d2)(e2) Only the normalized total light intensities (bucket intensities of BDs, fluctuating over time), $I_{AB}$ and $I_{BA}$, are accessible in the data post-processing.}
\label{figa}
\end{figure}

In the correlator, a He-Ne laser (632.8 nm) collimated by a telescope is split into two beams by a 50/50 BS (not shown here), and the two beams pass through two different rotating ground glass plates (RGGPs) to form statistically independent pseudo-thermal light sources, A and B. The pseudo-thermal light beam A (B) is then divided into two beams, $A_x$($B_x$) and $A_y$ ($B_y$), through BS$_1$ (BS$_2$). Obviously, the two sub-beams $A_x$ and $B_y$ ($A_y$ and $B_x$) are uncorrelated by nature. Refer to Methods for experimental details.
Figures~\ref{figa}(a1)(a2) show the instantaneous intensity profiles \(I_x\) (\(I_y\)) of the two independent pseudo-thermal beams, and Figs.~\ref{figa}(b1)(b2) and (c1)(c2) show the normalized auto- and cross-correlation functions \(g^{(2)}\), respectively.
The calculation of the normalized intensity cross-correlation function can be expressed as:
\begin{equation}
g^{(2)}\left[I_x(\mathbf{r}_x),I_y(\mathbf{r}_y);\tau=0\right]=\frac{\left \langle I_x(\mathbf{r}_x)I_y(\mathbf{r}_y)\right \rangle}{\left \langle I_x(\mathbf{r}_x)\right \rangle \left \langle I_y(\mathbf{r}_y)\right \rangle},
\label{eq:0}
\end{equation}
where $\left \langle \cdot \right \rangle$ denotes the ensemble average over time, $\mathbf{r}\equiv(x,y)$ is the transverse position, and this work does not consider the time domain, so the time difference $\tau$ defaults to 0. In this paper, the observation method involved fixing a point (i.e., midpoint) in one detector and performing joint measurements with a region of interest in another detector to obtain a two-dimensional distribution of $g^{(2)}(\mathbf{r})$ over 750k realizations or shots.

The results in Fig.~\ref{figa}(c1)(c2) confirm the absence of spatial correlation between the two laboratory-prepared pseudo-thermal light sources.
To generate a certain intensity correlation between the spatially separated and uncorrelated beams $A_x$ and $B_y$, as shown in Fig.~\ref{fig1}, we mix the two sub-beams $A_y$ and $B_x$ in BS$_3$ and obtain two outgoing intensity distributions $I_{ab}(\mathbf{r})$ and $I_{ba}(\mathbf{r})$.
This core component of the proposed classical correlator resembles the Mach-Zehnder interferometer. 
It should be noted that we only measure the total intensity here, i.e., perform bucket detection as shown in Fig.~\ref{figa}(d2)(e2), at the two output ports of BS$_3$, via
\begin{equation}
I_{AB}=\sum_{i=1}^{N}I_{ab}(\mathbf{r}_{i}),\,\,\,
I_{BA}=\sum_{i=1}^{N}I_{ba}(\mathbf{r}_{i}),
\label{eq:00}
\end{equation}
where $N$ is the total number of grid points (or pixels) in the region of interest on the detector. To facilitate the subsequent theoretical derivation, we let the size of each grid cell be equivalent to the average size of the speckle grain in Figs.~\ref{figa}(d1)(e1). 
In the experimental situation, the value of \(N\) is approximately 56, which means that there are on average 69 independent spatial speckle modes within the field of view.
Like other GI schemes \cite{XuYK2015,YeZ2020}, this correlator is immune to strong scattering because it only requires the temporally fluctuating bucket intensities rather than spatial distributions. Hence, this property allows, for instance, a diffuser to be placed directly in front of the BDs, significantly relaxing the alignment precision requirement for the interferometer and thereby enhancing its practicality.
Since the detectors placed at the two output ports of BS$_3$ are spatially unresolved, this interferometer can be regarded as a spatially unresolved Mach-Zehnder interferometer.
Based on this setup in Fig.~\ref{fig1}, we next introduce two data post-processing schemes to generate and even manipulate the transverse correlations between $A_x$ and $B_y$.

 \subsection{Scheme I: medium variable}
 \label{sec:2:2}
 In the first scheme, with the two bucket signals ($I_{AB}$ and $I_{BA}$), we define a new random variable as a mediating variable
\begin{equation}
C \equiv \frac{(I_{AB}-I_{BA})^2}{\langle (I_{AB}-I_{BA})^2 \rangle},
\label{eq:1}
\end{equation}
which is a normalized random variable \textit{without spatial resolution}. Obviously, the variable $C$ includes the intensity information of both input beams $A$ and $B$. For a 50/50 BS$_3$, the two input fields ($E_{Ay}$,$E_{Bx}$) and two output fields ($E_{ab}$,$E_{ba}$) satisfy
\begin{eqnarray}
E_{ab}=\frac{1}{\sqrt{2}} (E_{Ay}+E_{Bx}),\quad
E_{ba}=\frac{1}{\sqrt{2}} (E_{Ay}-E_{Bx}).
\label{eq:2}
\end{eqnarray}
The corresponding intensity distributions are given by
\begin{equation}
\begin{aligned}
I_{ab}(\mathbf{r})
&=\frac{1}{2}\left[I_{x}(\mathbf{r})+I_{y}(\mathbf{r})
+2\sqrt{I_{x}(\mathbf{r})I_{y}(\mathbf{r})}\cos{\varphi}(\mathbf{r})\right],\\
I_{ba}(\mathbf{r})
&=\frac{1}{2}\left[I_{x}(\mathbf{r})+I_{y}(\mathbf{r})
-2\sqrt{I_{x}(\mathbf{r})I_{y}(\mathbf{r})}\cos{\varphi}(\mathbf{r})\right].
\end{aligned}
\label{eq:3}
\end{equation}
where $\varphi$ is the phase difference between the two input fields $E_{Ay}$ and $E_{Bx}$. Without losing generality, let us assume that both BS$_1$ and BS$_2$ are 50/50 BS, so $I_x=|E_{Ay}|^2$ and $I_y=\left|E_{Bx}\right|^2$. Using Eq.~(\ref{eq:3}), we can calculate
\begin{equation}
I_{AB}-I_{BA}
=2\sum_{i=1}^{N}\sqrt{I_{x}(\mathbf{r}_i)I_{y}(\mathbf{r}_i)}\cos{\varphi}(\mathbf{r}_i)
\label{eq:4}
\end{equation}
and
\begin{equation}
\begin{aligned}
\left(I_{AB}-I_{BA}\right)^2
={}&\,4\sum_{i=1}^{N} I_{x}(\mathbf{r}_i) I_{y}(\mathbf{r}_i)\cos^2\!\varphi(\mathbf{r}_i) \\
&+ 8\sum_{i,j=1\,(i\neq j)}^{N}
\sqrt{
I_{x}(\mathbf{r}_{i}) I_{y}(\mathbf{r}_{i})
I_{x}(\mathbf{r}_{j}) I_{y}(\mathbf{r}_{j})
}
\,\cos\!\varphi(\mathbf{r}_{i})\cos\!\varphi(\mathbf{r}_{j}) .
\end{aligned}
\label{eq:5}
\end{equation}

Taking into account that all speckle grains are statistically independent of each other, the intensities and phases of each input thermal beam are also statistically independent, and the phase difference $\varphi(\mathbf{r}_{i})$ at each realization is an independent random variable with equal probability in $[0, 2\pi]$, we obtain
\begin{equation}
\begin{aligned}
\Bigl\langle\left(I_{AB}-I_{BA}\right)^2\Bigr\rangle
&=4\Bigl\langle\sum_{i=1}^{N}I_{x}(\mathbf{r}_i)I_{y}(\mathbf{r}_i)\cos^2{\varphi}(\mathbf{r}_i)\Bigr\rangle 
=4N\left\langle I_{x}\right\rangle\left\langle I_{y}\right\rangle\left\langle \cos^2{\varphi}\right\rangle .
\end{aligned}
\label{eq:6}
\end{equation}
We now consider the correlation of the medium $C$ with beams $A$ and $B$, and first calculate
\begin{equation}
\begin{aligned}
&\left\langle I_{x}(\mathbf{r}_{x})(I_{AB}-I_{BA})^{2}\right\rangle
= {} 4\Bigl\langle I_{x}(\mathbf{r}_{x})
\Bigl[\sum_{i=1}^{N}I_{x}(\mathbf{r}_{i})I_{y}(\mathbf{r}_{i})\cos^{2}\varphi(\mathbf{r}_{i})\Bigr]\Bigr\rangle \\
= {}& 4\Bigl\langle I_{x}(\mathbf{r}_{x})
\Bigl[\sum_{i=1,i\neq x}^{N-1}I_{x}(\mathbf{r}_{i})I_{y}(\mathbf{r}_{i})\cos^{2}\varphi(\mathbf{r}_{i})+I_{x}(\mathbf{r}_{x})I_{y}(\mathbf{r}_{x})\cos^{2}\varphi(\mathbf{r}_{x})\Bigr]\Bigr\rangle \\
= {}& 4(N-1)\left\langle I_{x}\right\rangle^{2}\left\langle I_{y}\right\rangle
\left\langle\cos^{2}\varphi\right\rangle+4\left\langle I_{x}^{2}\right\rangle\left\langle I_{y}\right\rangle
\left\langle\cos^{2}\varphi\right\rangle \\
= {}& 4\left\langle I_{y}\right\rangle\left\langle\cos^{2}\varphi\right\rangle
\left[N\left\langle I_{x}\right\rangle^{2}
+\left\langle I_{x}^{2}\right\rangle-\left\langle I_{x}\right\rangle^{2}\right].
\end{aligned}
\label{eq:7}
\end{equation}
Based on Eqs.~(\ref{eq:6})(\ref{eq:7}), the normalized correlation between $I_x(\mathbf{r}_x)$ and $C$ can be written as
\begin{equation}
\begin{aligned}
\frac{\left\langle I_x(\mathbf{r}_x)C\right\rangle}
{\left\langle I_x\right\rangle\left\langle C\right\rangle}=
\frac{\Bigl\langle I_x(\mathbf{r}_x)(I_{AB}-I_{BA})^2\Bigr\rangle}
{\left\langle I_x\right\rangle\Bigl\langle\left(I_{AB}-I_{BA}\right)^2\Bigr\rangle} =
1+\frac{1}{N}\frac{\left\langle I_x^2\right\rangle-\left\langle I_x\right\rangle^2}
{\left\langle I_x\right\rangle^2} =
1+\frac{1}{N}\left({g}_x^{(2)}-1\right).
\end{aligned}
\label{eq:8}
\end{equation}
where $g^{(2)}_x \equiv \langle I_x^2\rangle/\langle I_x\rangle^2$ ($g^{(2)}_y \equiv \langle I_y^2\rangle/\langle I_y\rangle^2$) is the normalized intensity auto-correlation of beam A (B). 
Similarly, we have
\begin{equation}
\begin{aligned}
\frac{\left\langle I_y(\mathbf{r}_y)C\right\rangle}
{\left\langle I_y\right\rangle\left\langle C\right\rangle}=
1+\frac{1}{N}\frac{\left\langle I_y^2\right\rangle-\left\langle I_y\right\rangle^2}
{\left\langle I_y\right\rangle^2} =
1+\frac{1}{N}\left(g_y^{(2)}-1\right).
\end{aligned}
\label{eq:9}
\end{equation}

The above results indicate that the random variable $C$ can form a weak correlation, inversely proportional to $N$, with two independent thermal light beams individually. However, this correlation does not produce distinct peaks [such as in Figs.~\ref{figa}(b1)(b2)], resembling a background slightly greater than 1 for all positions $\mathbf{r}_x$ or $\mathbf{r}_y$. Our strategy is to use $C$ as a medium variable to connect two uncorrelated beams. We calculate the intensity correlation of three random variables, akin to the form of the {third-order correlation function}, for the two cases, $\mathbf{r}_x\neq\mathbf{r}_y$ and $\mathbf{r}_x = \mathbf{r}_y$, as follows:

(i) $\mathbf{r}_x\neq\mathbf{r}_y$
\begin{equation}
\begin{aligned}
&\left\langle I_x(\mathbf{r}_x)I_y(\mathbf{r}_y)(I_{AB}-I_{BA})^2\right\rangle \\
&=4\Bigl\langle I_x(\mathbf{r}_x)I_y(\mathbf{r}_y)
\Bigl[\sum_{i=1}^N I_x(\mathbf{r}_i)I_y(\mathbf{r}_i)\cos^2\varphi(\mathbf{r}_i)\Bigr]\Bigr\rangle \\
&=4\Bigl\langle I_x(\mathbf{r}_x)I_y(\mathbf{r}_y)\Bigl[
\sum_{i\neq x,y}^{N-2}I_x(\mathbf{r}_i)I_y(\mathbf{r}_i)\cos^2\varphi(\mathbf{r}_i)
+I_x(\mathbf{r}_x)I_y(\mathbf{r}_x)\cos^2\varphi(\mathbf{r}_x)\\
&\qquad\qquad\qquad\qquad
+I_x(\mathbf{r}_y)I_y(\mathbf{r}_y)\cos^2\varphi(\mathbf{r}_y)\Bigr]\Bigr\rangle \\
&=4\left\langle\cos^2{\varphi}\right\rangle
\left[
N\left\langle I_x\right\rangle^2\left\langle I_y\right\rangle^2
+\left(\left\langle I_x^2\right\rangle-\left\langle I_x\right\rangle^2\right)\left\langle I_y\right\rangle^2
+\left(\left\langle I_y^2\right\rangle-\left\langle I_y\right\rangle^2\right)\left\langle I_x\right\rangle^2
\right],
\end{aligned}
\label{eq:10}
\end{equation}

\begin{equation}
\begin{aligned}
&\frac{\left\langle I_{x}(\mathbf{r}_{x})I_{y}(\mathbf{r}_{y})C\right\rangle}
{\left\langle I_{x}\right\rangle\left\langle I_{y}\right\rangle\left\langle C\right\rangle}
=
\frac{\left\langle I_{x}(\mathbf{r}_{x})I_{y}(\mathbf{r}_{y})(I_{AB}-I_{BA})^{2}\right\rangle}
{\left\langle I_{x}\right\rangle\left\langle I_{y}\right\rangle\left\langle(I_{AB}-I_{BA})^{2}\right\rangle} \\
&=
1+\frac{1}{N}\left[
\frac{\left\langle I_{x}^{2}\right\rangle-\left\langle I_{x}\right\rangle^{2}}{\left\langle I_{x}\right\rangle^{2}}
+\frac{\left\langle I_{y}^{2}\right\rangle-\left\langle I_{y}\right\rangle^{2}}{\left\langle I_{y}\right\rangle^{2}}
\right] \\
&=
1+\frac{1}{N}\left[g_{x}^{(2)}+g_{y}^{(2)}-2\right],
\end{aligned}
\label{eq:11}
\end{equation}

(ii) $\mathbf{r}_x=\mathbf{r}_y$
\begin{equation}
\begin{aligned}
&\left\langle I_x(\mathbf{r}_x)I_y(\mathbf{r}_x)(I_{AB}-I_{BA})^2\right\rangle
=4\Bigl\langle I_x(\mathbf{r}_x)I_y(\mathbf{r}_x)
\Bigl[\sum_{i=1}^N I_x(\mathbf{r}_i)I_y(\mathbf{r}_i)\cos^2\varphi(\mathbf{r}_i)\Bigr]\Bigr\rangle \\
&=4\Bigl\langle I_{x}(\mathbf{r}_{x})I_{y}(\mathbf{r}_{y})
\Bigl[\sum_{i\neq x}^{N-1}I_{x}(\mathbf{r}_{i})I_{y}(\mathbf{r}_{i})\cos^{2}\varphi(\mathbf{r}_{i})
+I_{x}(\mathbf{r}_{x})I_{y}(\mathbf{r}_{x})\cos^{2}\varphi(\mathbf{r}_{x})\Bigr]\Bigr\rangle \\
&=4\left\langle\cos^2{\varphi}\right\rangle
\left[
N\left\langle I_x\right\rangle^2\left\langle I_y\right\rangle^2
+\left\langle I_x^2\right\rangle\left\langle I_y^2\right\rangle
-\left\langle I_x\right\rangle^2\left\langle I_y\right\rangle^2
\right].
\end{aligned}
\label{eq:12}
\end{equation}

\begin{equation}
\begin{aligned}
\frac{\left\langle I_x(\mathbf{r}_x)I_y(\mathbf{r}_x)C\right\rangle}
{\left\langle I_x\right\rangle\left\langle I_y\right\rangle\left\langle C\right\rangle}
=
1+\frac{1}{N}\left[
\frac{\left\langle I_x^2\right\rangle\left\langle I_y^2\right\rangle}
{\left\langle I_x\right\rangle^2\left\langle I_y\right\rangle^2}-1
\right] 
=
1+\frac{1}{N}\left[g_x^{(2)}g_y^{(2)}-1\right].
\end{aligned}
\label{eq:13}
\end{equation}
Their difference \(\Delta g^{(2)}\) is obtained to be
\begin{equation}
\begin{aligned}
\Delta g^{(2)}=\frac{\langle I_x(\mathbf{r}_x)I_y(\mathbf{r}_x)C\rangle
-\langle I_x(\mathbf{r}_x)I_y(\mathbf{r}_y)C\rangle}
{\langle I_x\rangle\langle I_y\rangle\langle C\rangle}
=
\frac{1}{N}(g_x^{(2)}-1)(g_y^{(2)}-1).
\end{aligned}
\label{eq:14}
\end{equation}
Therefore, the medium variable $C$ as well as the {third-order correlation function} establishes the intensity correlation between the corresponding positions of two uncorrelated thermal light beams A and B, which still constitutes a very weak correlation. This manifests as a faint correlation peak superimposed on a background slightly above 1.

\begin{figure}[htbp]
\centering
{\includegraphics[width=\linewidth]{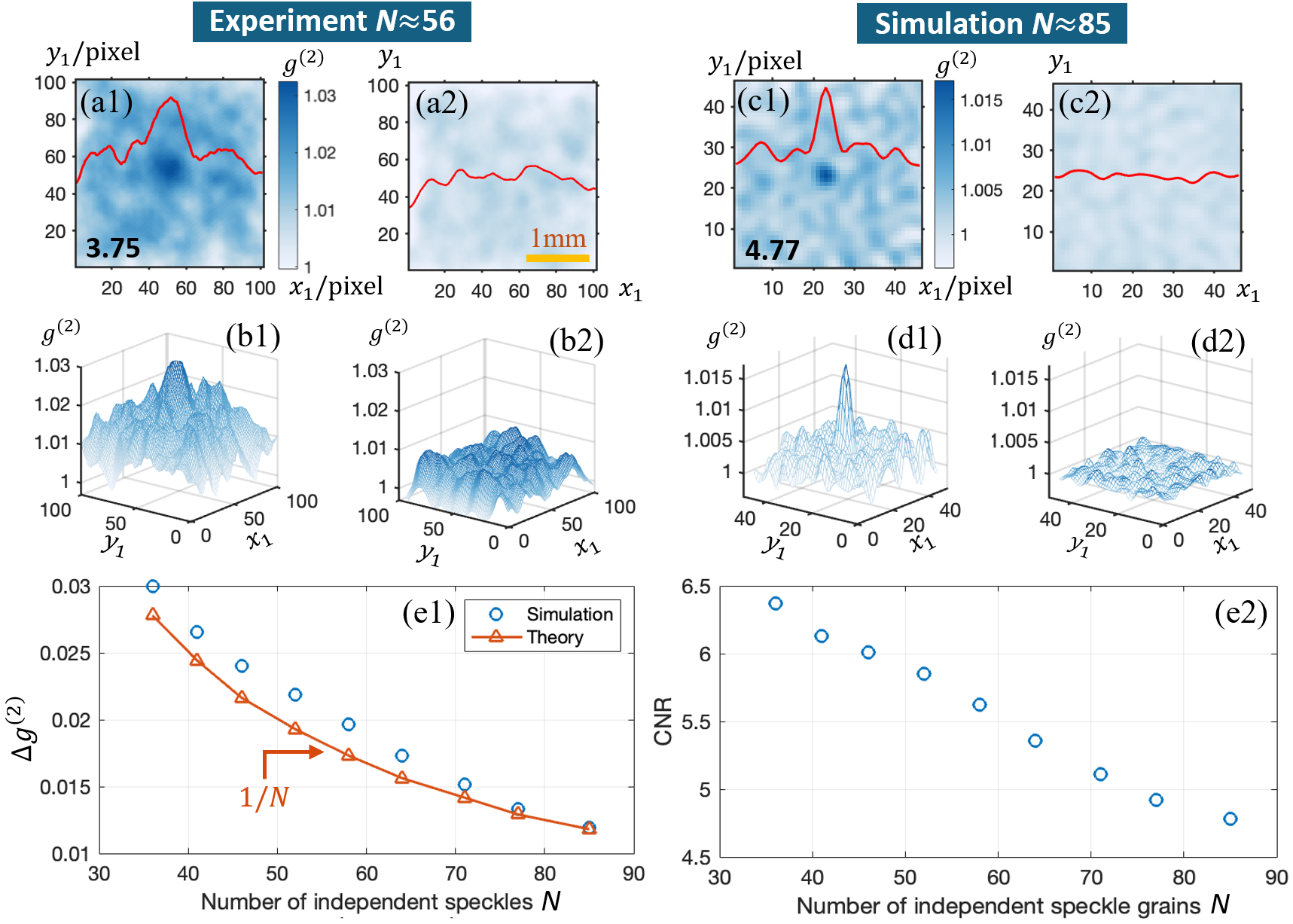}}
\caption{Experimental and simulation results of the correlator with scheme I.
(a1-b2) Experimental observations of generated faint correlation between two independent pseudo-thermal light beams through a medium variable $C$ ($N\approx56$). (c1-d2) Numerical results with $N\approx85$.
(a1)(c1) Spatial profile of normalized intensity correlation function, $\left<I_x\left ( \mathbf{r}_x \right )I_y\left ( \mathbf{r}_y \right )C \right>/\left ( \left<I_x \right>\left<I_y \right>\left<C \right> \right )$ when $\mathbf{r}_x$ is scanned and $\mathbf{r}_y$ is fixed at the midpoint. (a2)(c2) Normalized correlation function, $\langle I_x(\mathbf{r}_x)C\rangle/(\langle I_x\rangle\langle C\rangle)$. The corresponding 3D profiles are shown in (b1)(b2) and (d1)(d2), respectively.
(e1)(e2) \(\Delta g^{(2)}\) and CNR curves as functions of \(N\) in the numerical simulation. The numbers in the lower left corners of panels (a1) and (c1) are the CNR values.
}
\label{fig:2}
\end{figure}

By the experimental setup in Fig.~\ref{fig1}, we measure the normalized intensity correlation distribution $\langle I_x(\mathbf{r}_x)I_y(\mathbf{r}_y)C\rangle/(\langle I_x\rangle\langle I_y\rangle\langle C\rangle)$ when $\mathbf{r}_x$ is scanned and $\mathbf{r}_y$ is fixed at the midpoint. After \(M=\)750k shots, the experimental results of the 2D and 3D plots are shown in Figs.~\ref{fig:2}(a1) and (b1), respectively. 
By contrast to Figs.~\ref{figa}(c1)(c2), the intensity correlation peak occurs at the fixed position $\mathbf{r}_y$ with the value of 1.033, against the background distribution in the range of $1.017(4)$. The correlation effect is still valid for the experimental case of \(N\approx56\), although it is quite weak. The contrast-to-noise ratio (CNR) of the obtained correlation pattern is defined as $|\mu_s-\mu_b|/\sqrt{\sigma_s^2+\sigma_b^2}$ \cite{Chan2010}, where $\mu_\mathrm{s}$ and $\mu_b$ are mean values, $\sigma_\mathrm{s}$ and $\sigma_b$ are standard deviations for signal and background, respectively. When the CNR of the correlation pattern is less than 1, the signal is completely unrecognizable in the background. We obtain the CNR of 3.75 and the full width at half maximum (FWHM) of 526 $\mu$m for the correlation signal. For comparison, the normalized intensity correlation distribution $\langle I_x(\mathbf{r}_x)C\rangle/(\langle I_x\rangle\langle C\rangle)$ is shown in Figs.~\ref{fig:2}(a2)(b2). Likewise, Figs.~\ref{fig:2}(c1)-(d2) display the numerical results of \(N\approx85\) over \(M=\)2000k realizations (refer to the Method section for details). 
In the numerical simulation, we also calculated the curves of \(\Delta g^{(2)}\) and CNR as functions of \(N\) as shown in Figs.~\ref{fig:2}(e1)(e2), where the variation trend of \(\Delta g^{(2)}\) is approximately inversely proportional to \(N\), verifying the conclusion in Eq.~(\ref{eq:14}).

\subsection{Scheme II: post-selection}
\label{sec:2:3}
 
The previous scheme introduces a medium variable $C$ and implements correlation swapping by a third-order correlation function.
According to Eq.~(\ref{eq:14}), the generated correlation of two independent beams by the medium $C$ is inversely proportional to the total number $N$ of speckle grains. Since the visibility of the correlation signal is quite low for a large $N$, the observation of the correlation effect requires a significant increase in the number of realizations $M$. Hence, a huge amount of statistical data needs to be processed. To reduce the amount of statistical data and improve measurement efficiency, we propose a post-selection scheme with conditional correlation functions.

As shown in Fig.~\ref{figa}, the two output intensities $I_x(\mathbf{r}_x)$ and $I_y(\mathbf{r}_y)$ do not correlate with each other. However, if we select their data according to certain rules, the correlation may appear. To do this, we compose another medium random variable by the intensity difference $\Delta I=I_{AB}-I_{BA}$, which fluctuates around 0 with $\langle \Delta I\rangle=0$. Then we define the {conditional intensity correlation function} $\left\{ g^{(2)}\left[I_x(\mathbf{r}_x),I_y(\mathbf{r}_y)\right]\,|\,|\Delta I|\leq s\sigma\right\}$, in which the second-order intensity correlation function $g^{(2)}\left[I_x(\mathbf{r}_x),I_y(\mathbf{r}_y)\right]$ are selected by the given condition $|\Delta I|\leq s\sigma$ (or $|\Delta I|> s\sigma$) being satisfied, and $\sigma$ is the standard deviation of the random variable $\Delta I$. The scheme is similar to the conditional-averaging GI or correspondence imaging \cite{Luo2012,LiMF2013,Yang2018,LengJ2020,HuC2021,CaoDZ2024}, where the condition comes from the correlated variable itself.
There is a subtle difference between the two: correspondence imaging uses the bucket signal as a threshold to conditionally average the intensity profile of the optical field in the reference arm, whereas the scheme here uses the intensity difference \(\Delta I\) between the two bucket signals as a threshold to conditionally average the intensity correlation function.

\begin{figure*}[htbp]
\centering
{\includegraphics[width=\linewidth]{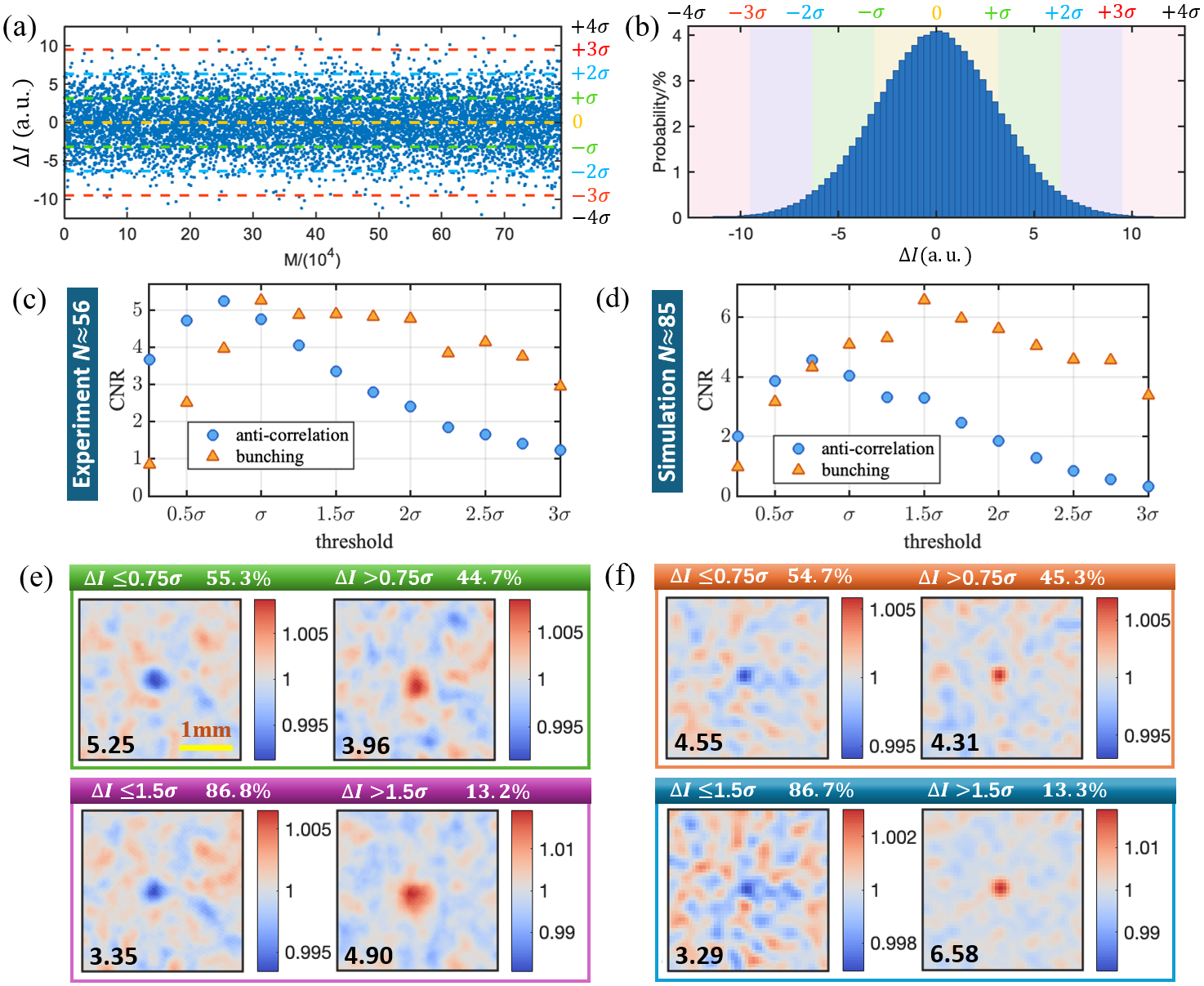}}
\caption{Experimental and numerical results of the conditional-averaging intensity correlation of two independent pseudo-thermal light sources (Scheme II). (a) Intensity evolution of $\Delta I$ in 750k shots, one point is shown in the plot for every 250 shots. (b) The probability distribution of $\Delta I$, where $\sigma$ is the standard deviation. (c) and (d) are the experimental and simulated curves of CNR as a function of the threshold $s\sigma$, respectively, for the two types of photon correlation produced in Scheme II. (e) and (f) show the experimental and simulated spatial distributions of the conditional correlation functions at two thresholds, respectively, where the number in percentage is the ratio of the total number of measurements. The numbers in the lower left corners of panels (e) and (f) are the CNR values.
Refer to Fig.~S1 and Fig.~S2 for complete experimental and simulated data.
}
\label{fig3}
\end{figure*}

In the experiment, after 750k shots of the random signals $\Delta I$, the evolution and corresponding probability distribution are shown in Figs.~\ref{fig3}(a) and (b), respectively. According to all recorded differential signals, we choose 12 values of $s\sigma$ as boundaries and evaluate the conditional intensity correlation distributions $\left\{ g^{(2)}\left[I_x(\mathbf{r}_x),I_y(\mathbf{r}_y)\right]\,|\,|\Delta I|\leq s\sigma\right\}$ and $\left\{ g^{(2)}\left[I_x(\mathbf{r}_x),I_y(\mathbf{r}_y)\right]\,|\,|\Delta I|>s\sigma\right\}$ when $\mathbf{r}_x$ is scanned and $\mathbf{r_y}$ is fixed at the midpoint. All the experimental results are displayed in Fig.~S1 and Fig. S2. 
For the cases of $|\Delta I|\leq s\sigma$, anti-bunching-like correlation dips $({g}^{(2)}<1)$ are observed, while bunching-type correlation peaks $({g}^{(2)}>1)$ occur for conditions of $|\Delta I|>s\sigma$. These behaviors are similar to the case of correspondence imaging, where a negative (or positive) image appears when the bucket signal is less (or greater) than the mean value \cite{Luo2012,LiMF2013,Yang2018,LengJ2020,HuC2021,CaoDZ2024}.
Figures~\ref{fig3}(c) and (d) respectively plot the experimental and simulated curves of CNR as a function of the threshold $s\sigma$ for the two types of photon correlation produced in Scheme II.
Figures~\ref{fig3}(e) and (f) respectively show the experimental and simulated conditional correlation functions at two thresholds, where $\mathbf{r}_x$ is scanned and $\mathbf{r}_y$ is fixed at the midpoint. The number in percentage is the ratio of the total number of measurements.

Compared to Scheme I, this scheme can not only achieve correlation swapping but also flexibly manipulate the ``polarity'' of the induced spatial correlation, i.e., whether it exhibits bunching or anti-correlation. Moreover, by employing conditional averaging instead of ensemble averaging, it allows the use of much less data to obtain spatial correlations with better quality.
For instance, under the condition of $|\Delta I|>1.5\sigma$, the experimental CNR of the correlation peak reaches 4.90 with only $13.2\%$ of the total data (i.e., $M\approx99$k shots). When $\Delta I\leq0.75\sigma$, both experiments and simulations produce the correlation dip with the best CNR; while $s$ between 1 and 2 is the threshold condition to obtain the best correlation peak.
It should be noted that the threshold conditions for obtaining the best CNR for the positive and negative correlation patterns do not coincide symmetrically. In addition, it is experimentally shown that the average FWHM of the anti-bunching correlation is 436 $\mu$m, which is indeed smaller than that of the bunching one with 521 $\mu$m. This might be attributed to the low visibility of the correlation pattern, thereby introducing systematic errors in the calculation.
In summary, selecting an appropriate threshold condition can significantly reduce the amount of data acquired and substantially alleviate the processing throughput and memory demands on the hardware. That said, a rigorously proven theoretical foundation for determining the optimal threshold condition in our experiments remains an open question.

\subsection{Interaction-free GI using unknown light}
\label{sec:2:4}
  
\begin{figure}[htbp]
\centering
{\includegraphics[width=\linewidth]{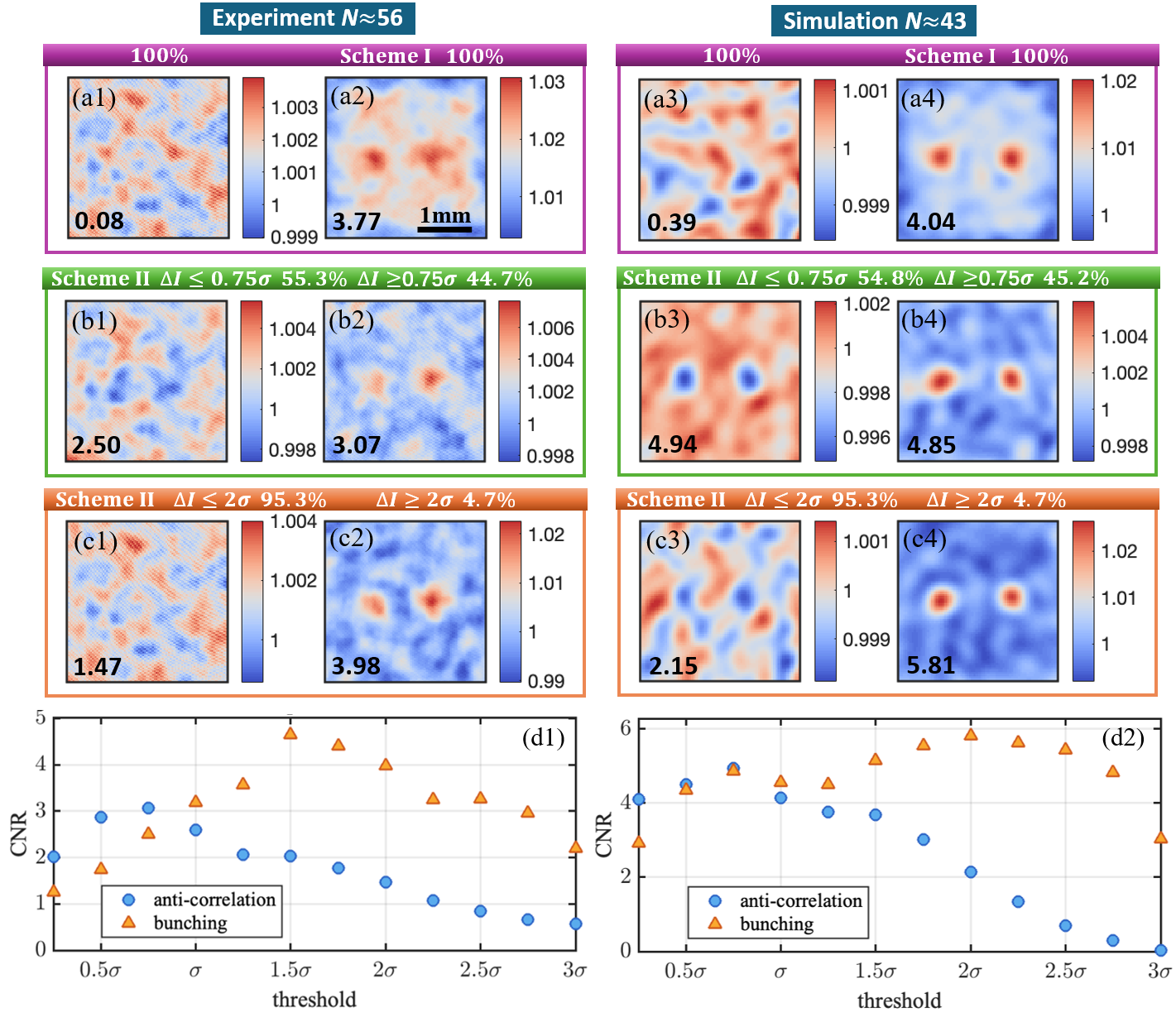}}
\caption{Experimental and numerical results of interaction-free GI experiments using uncorrelated light. (a1)(a3) GI without using the correlator. (a2)(a4) GI using correlator in Scheme I. (b)(c) GI using correlator in Scheme II for two boundaries $0.75\sigma$ and $2\sigma$. Full sampling: 750k shots (experiments, \(N\approx56\)) and 2000k shots (simulations, \(N\approx43\)). Sampling percentages and CNRs are also listed. 
(d) CNR curves of positive and negative ghost images as a function of threshold \(s\sigma\).
Also refer to Fig. S3 and Fig. S4 for the complete data.}
\label{fig:5}
\end{figure}

Using the correlator in Fig.~\ref{fig1}, we now perform interaction-free GI experiments with the two schemes. 
Here, `interaction-free' means that the object is illuminated by unknown or uncorrelated light, and no direct spatial correlation exists between the light that interacts with the object and the light used for imaging.
In this case, a two-pinhole object with a pinhole diameter $\sim$40 $\mu$m and a spacing $\sim$1 mm is placed on the plane of AD$_2$, and a BD is placed behind the object to record the total light intensity, namely $I_Y=\sum_{i=1}^{N}I_{y}(\mathbf{r}_{i})$.
When the correlator is turned off, no valid image information can be obtained between two independent sources shown in Fig.~\ref{fig:5}(a1). 
By contrast, when the correlator of Scheme I is turned on, a two-spot correlation pattern via $\langle I_x(\mathbf{r}_x)I_YC\rangle/\langle I_x \rangle\langle I_Y \rangle\langle C\rangle$ is yielded in Fig.~\ref{fig:5}(a2) with the CNR of 3.77, and 750k shots contribute to image quality.
Similarly, Figs.~\ref{fig:5}(a3) and (a4) show the numerical results of \(N\approx43\) with 2000k realizations.
Figures~\ref{fig:5}(b) and (c) display two sets of GI using the correlator of Scheme II for two boundaries, 0.75$\sigma$ and 2$\sigma$ (refer to Fig. S3 and Fig. S4 for complete data).
These results indicate that positive ghost images tend to exhibit a higher CNR. Figure~\ref{fig:5}(b1) actually shows the dark ghost image (\(0.75\sigma\)) with the highest visibility in the experimental observation (CNR=3.07), while the positive image on the opposite side (Fig.~\ref{fig:5}(b2)) shows a CNR of 2.50. 
Figures~\ref{fig:5}(c2) and (c4) show the positive ghost images (2$\sigma$, CNR=3.98 and 5.81) of better quality with only a sampling rate of $4.7\%$, while the opposite condition cannot retrieve the dark image. 
Figures~\ref{fig:5}(d1) and (d2) plot the experimental and numerical CNR curves of positive and negative ghost images as a function of threshold \(s\sigma\), respectively.
In general, compared to Scheme I, Scheme II is more valuable and practical because it can reduce the amount of data and improve image quality. Compared to the observations in Figs.~\ref{fig:5}(d1)(d2), the optimal threshold condition also seems to be closely related to the object itself.

Notably, this is the first experimental demonstration, to the best of our knowledge, that classical GI cannot rely on correlated light but utilizes unknown random light. The word ``unknown'' here is embodied in the fact that the light field or its copy interacting with the object is not spatially resolved, recorded, or reconstructed. Instead, only an extremely low-dimensional signal fed into a single-pixel correlator plays a crucial role in implementing correlation swapping between known and unknown light.
This work is essentially different from classical or computational GI \cite{Erkmen2010}, in which the former requires correlated stochastic light and the latter requires pre-calibration or prediction of a repeatable random light. 
Hence, this experimental case might provide a new path for spatially inaccessible light (e.g., true thermal light) to be better applied to various applications such as GI.
The newly created correlations are very weak at this stage, and thus require a large number of realizations or significantly reduce the \(N\) value (imaging resolution), and might be only suitable for some simple and sparse objects when considering GI applications. 

\subsection{Extension of the correlator to OAM variables}
\label{sec:2:5}
The preceding subsections have successfully demonstrated correlation swapping between two independent pseudo-thermal light sources, with the transverse correlation primarily established in position space \((\mathbf{r}_a,\mathbf{r}_b)\). An interesting question naturally arises: can the two proposed schemes be extended to other transverse degrees of freedom, such as OAM variables \((\ell_a,\ell_b)\) \cite{MairA2001,LeachJ2010,Maga2016}? In this subsection, we numerically show that OAM correlations can also be generated between two independent pseudo-thermal light sources via the same correlator and the two proposed schemes.  

\begin{figure}[htbp]
\centering
{\includegraphics[width=\linewidth]{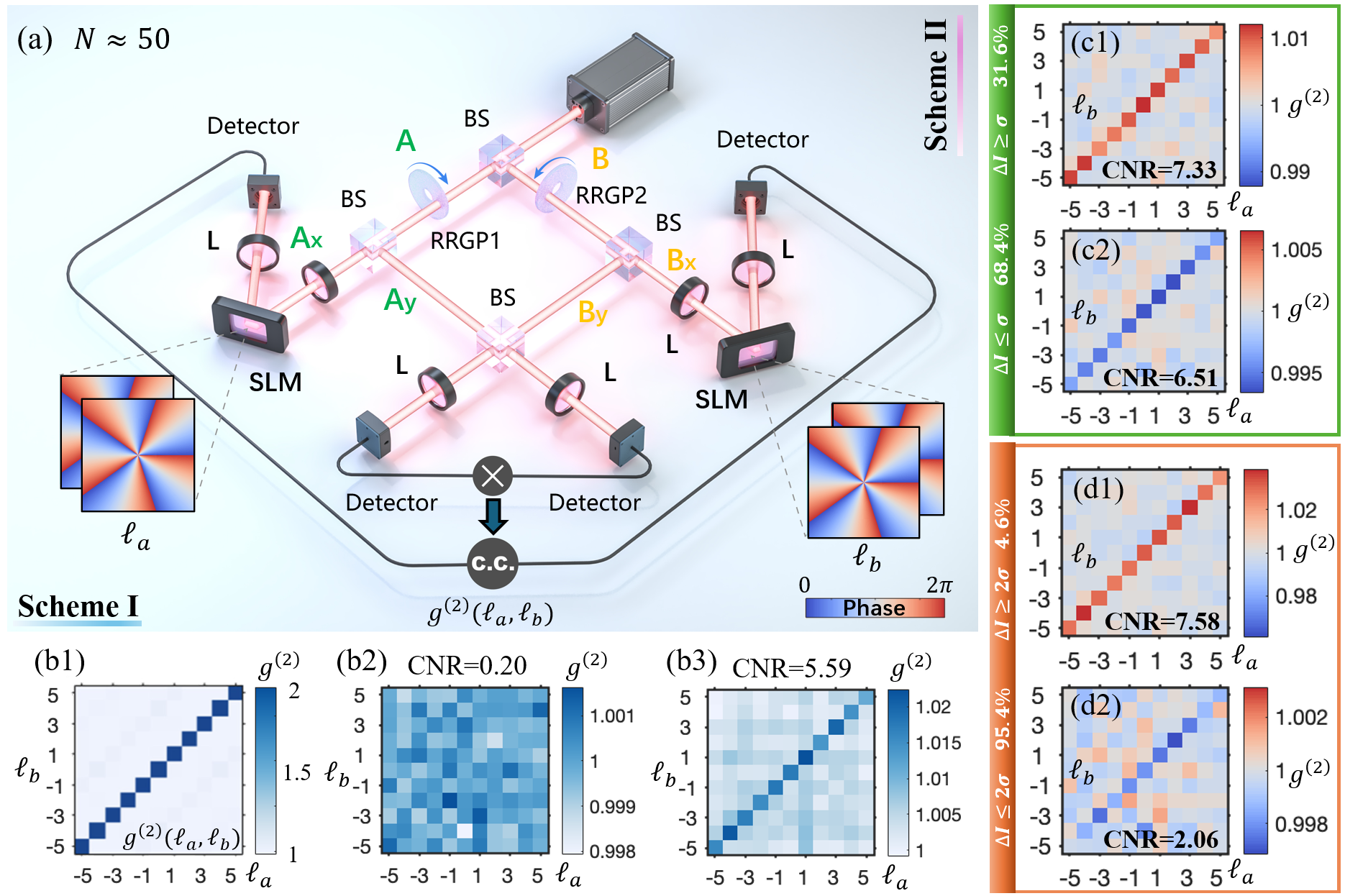}}
\caption{
Classical correlator for OAM variables using the two proposed schemes.
(a) Schematic diagram: the interferometer configuration is the same as that in Fig.~\ref{fig1}, and the difference is to measure the projections of $A_x$ and $B_y$ on 11 OAM modes, where $\ell_a$ and $\ell_b$ vary from $-5$ to $+5$. L: lens, SLM: spatial light modulator.
Numerical results of the OAM correlator of (b) Scheme I and (c)(d) Scheme II (\(N\approx50\)). (b1) shows the ideal OAM auto-correlation function \(g^{(2)}(\ell_a,\ell_b)\) in thermal light, where the correlation signals \(g^{(2)}\) of the anti‑diagonal elements satisfying \(\ell_a-\ell_b=0\) are greater than those of the other regions. (b2) and (b3) show the OAM correlation function between a pair of independent thermal light sources (without a correlation peak) and the OAM correlation function using the correlator based on Scheme I (with peak signals reappearing on the anti-diagonal), respectively. (c1,c2) and (d1,d2) are the OAM correlation functions obtained with the correlator of Scheme II under conditional averaging at different thresholds, respectively. 2000k realizations contribute to the quality of the OAM correlation function.
}
\label{fig:6}
\end{figure}

Figure~\ref{fig:6}(a) shows the setup of an OAM correlator, in which the detection port in Fig.~\ref{fig1} is replaced with an OAM-sensitive projection measurement: In the path of $A_x$ [$B_y$], the original AD in Fig.~\ref{fig1} is replaced with a spatial light modulator (SLM) imprinted with the helical phase profile with the OAM mode of $\ell_a$ [$\ell_b$], a lens (L) performs the optical Fourier transform, and the beam is then injected into a detector mounted with a single-mode fiber (SMF). The core ingredient, i.e., the spatially-unresolved Mach–Zehnder interferometer, is identical to Fig.~\ref{fig1}.
Furthermore, the observation plane in Fig.~\ref{fig1} is placed at a certain distance from the thermal source after free-space propagation, whereas in the OAM correlator here, the observation plane is set in the far field with \(N\approx50\), i.e. the optical Fourier transform plane (2f system). In this way, two SMF-mounted detectors measure the two intensity fluctuations, $I_x(\ell_a)$ and $I_y(\ell_b)$, from the OAM variables $\ell_a$ and $\ell_b$, respectively, and then calculate their correlation coefficient $g^{(2)}[I_y(\ell_b),I_x(\ell_a)]$ using Eq.~(\ref{eq:0}), where $\ell_a$ and $\ell_b$ are integers and take values from $-5$ to $+5$ of interest, i.e., 11 components. 
Refer to Methods and Section S3 of the supplementary information for more details.
As a result, $g^{(2)}[I_y(\ell_b),I_x(\ell_a)]$ here forms an 11$\times$11 OAM correlation function. According to Scheme I, the third-order correlation function can be expressed as 
${\left \langle I_y(\ell_b)I_x(\ell_a)C\right \rangle}/[{\left \langle I_y(\ell_b)\right \rangle \left \langle I_x(\ell_a)\right \rangle\left \langle C\right \rangle}]$, based on the medium variable $C$ in Eq.~(\ref{eq:1}).
As shown in Fig.~\ref{fig:6}(b1), a typical classical OAM autocorrelation function of thermal light \cite{Maga2016} shows bunching-type correlation peaks in the anti-diagonal area (\(\ell_a-\ell_b=0\)) against the background of $g^{(2)}=1$. 
For the two independent thermal light sources, no correlation peaks can be found in Fig.~\ref{fig:6}(b2) with CNR= \(0.20<1\), where anti-diagonal elements are completely submerged in noise. 
With Scheme I and the third-order correlation, numerical results in Fig.~\ref{fig:6}(b3) show that the positive (or bunching-type) OAM correlation is generated between two independent thermal light sources with a CNR of 5.59. Note that the CNR value here is calculated by treating the elements on the anti-diagonal region as the signal and the elements in all other regions as the background noise.

When Scheme II in Sec.~\ref{sec:2:3} is applied, we also define two conditional OAM correlation functions $\left\{ g^{(2)}\left[I_x(\ell_a),I_y(\ell_b)\right]\,|\,|\Delta I|\leq s\sigma\right\}$ and $\left\{ g^{(2)}\left[I_x(\ell_a),I_y(\ell_b)\right]\,|\,|\Delta I|>s\sigma\right\}$, based on the other medium variable $\Delta I$.
As shown in Figs.~\ref{fig:6}(c)(d), similarly to Figs.~\ref{fig3}(e)(f), we show two types of engineered OAM correlation functions at two different thresholds (\(\sigma\) and \(2\sigma\)), in which a novel classical OAM correlation mode—anti-bunching-like OAM correlation dips with the anti-diagonal's correlation coefficients less than the background—can be realized for the case of $\Delta I\leq s\sigma$. 
Intriguingly, such unique OAM correlation dips have not yet been reported in classical light, but in some quantum interference experiments, such as Hong-Ou-Mandel interference in OAM variables \cite{ZhangY2016,LiuZF2022,HongL2023}. 
The thresholds for obtaining the high-CNR OAM correlation dip and peak are not coincident, as in the case in Fig.~\ref{fig3}. Moreover, a sampling rate as low as 4.6$\%$ (refer to Fig.~\ref{fig:6}(d1)) can yield considerable OAM correlations that might potentially be utilized in angular object identification \cite{YangZ2017,YeZy2023} and many OAM-based classical information transmission tasks \cite{Qiu2023,GongL2019,Ma2024,YeYz2025} from nonlocal to interaction-free \cite{ZHangY2019,YangY2023}.

\subsection{Correlator for thermal light with different types of spatial correlations}
\label{sec:2:6}
We have already demonstrated that the classical correlator, along with two signal processing schemes, can induce transverse correlations between two independent thermal light sources in at least two spatial degrees of freedom (position/momentum and OAM).
In this subsection, we will finally demonstrate that the correlator can even establish correlations between two thermal light sources that are not only independent but also of different spatial correlation types.

Generally, there are two types of spatial correlation \cite{Bennink2004,Gatti2004,Brambilla2004,Magana2019a}: one is the HBT correlation present in thermal light, and the other is the Einstein-Podolsky-Rosen (EPR) correlation present in entangled photon pairs. Recently, we have fabricated a novel pseudo-thermal light source called holographic thermal light \cite{YEZY2025,YE2026}, which can generate pairs of conjugate chaotic beams exhibiting EPR-like spatial correlations.
This pair of conjugate chaotic beams exhibits an evolution from (angular) position correlation in the near field to (orbital angular) momentum anti-correlation in the far field.
The preceding results have demonstrated the effectiveness of this correlator for conventional HBT-type thermal light. A natural question is whether this correlator is applicable to holographic thermal light with EPR-like correlations, or even to two types of thermal light with distinctly different correlation properties.

\begin{figure}[htbp]
\centering
{\includegraphics[width=\linewidth]{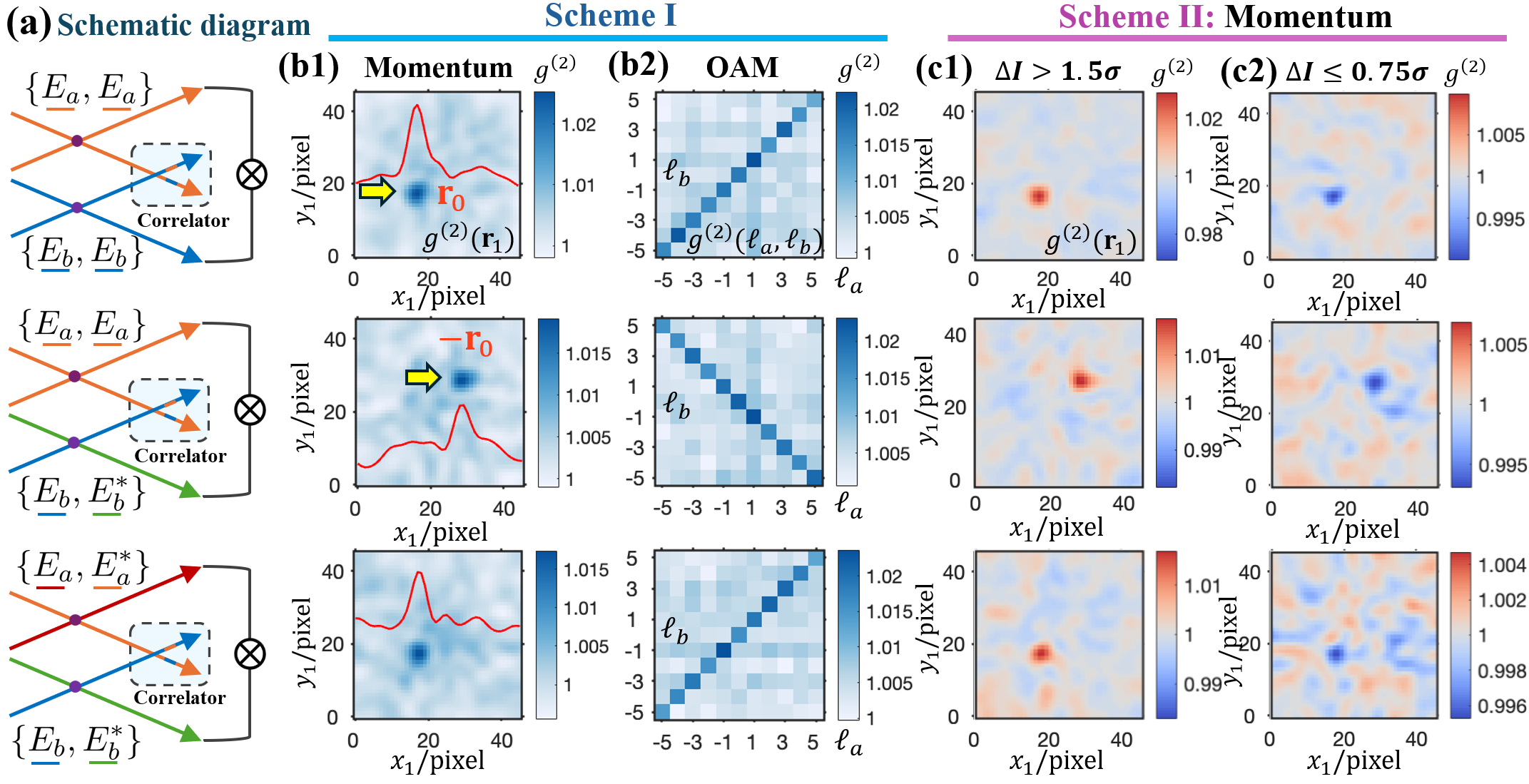}}
\caption{Correlator for two independent thermal light sources with different types of spatial correlations (HBT and EPR-like). (a) Schematic diagram of three cases (from top to bottom): HBT\(\oplus\)HBT, HBT\(\oplus\)EPR-like, EPR-like\(\oplus\)EPR-like. (b1) and (b2) Numerical results of induced momentum and OAM correlation functions of the three cases using the correlator (Scheme I). (c1) and (c2) Numerical results of induced momentum correlation functions of the three cases using the correlator (Scheme II) under two threshold conditions (\(0.75\sigma\) and \(1.5\sigma\)).
}
\label{fig:7}
\end{figure}

Figure~\ref{fig:7}(a) shows the schematic diagram of three cases: From top to bottom, they are two independent HBT-type thermal lights (HBT\(\oplus\)HBT, this case has been fully studied in the previous subsections), HBT-type thermal light and holographic thermal light (HBT\(\oplus\)EPR), and two independent holographic thermal lights (or two independent pairs of conjugate chaotic beams, EPR\(\oplus\)EPR).
In this case, the numerical simulation parameters are the same as those in Fig.~\ref{fig:6} (\(N\approx50\)). In a simplified mathematical model, HBT-type thermal light can be written as a pair of identical chaotic light fields (e.g., \(\{E_a,E_a\}\) and \(\{E_b,E_b\}\)), while holographic thermal light can be written as a pair of conjugate chaotic light fields (e.g., \(\{E_a,E^*_a\}\) and \(\{E_b,E^*_b\}\)). The observation plane is set in the far field, and correlations in both the momentum and OAM spatial degrees of freedom are considered.

\begin{figure}[htbp]
\centering
{\includegraphics[width=0.85\linewidth]{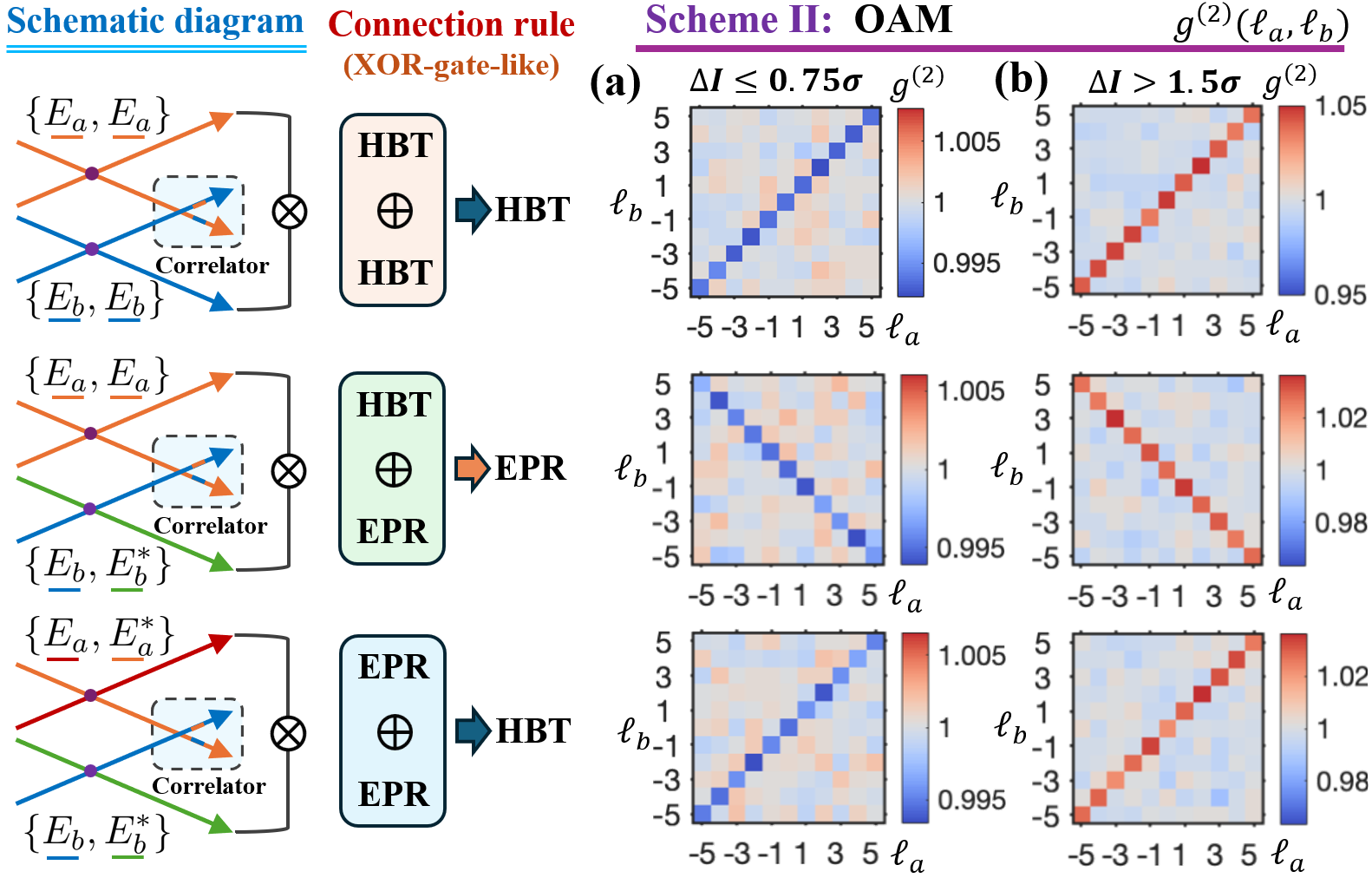}}
\caption{Numerical results of OAM correlator (Scheme II) for two independent thermal light sources with different types of spatial correlations. (a) and (b) show the OAM correlation functions \(g^{(2)}(\ell_a,\ell_b)\) under two threshold conditions, respectively.
}
\label{fig:8}
\end{figure}

We discover some interesting connection rules (refer to Sections S2 and S3 in the supplementary information for detailed theoretical derivation). Between independent thermal lights with the same correlation properties, an HBT-type correlation is established (HBT\(\oplus\)HBT, EPR\(\oplus\)EPR \(\rightarrow\) HBT), whereas between those with opposite correlation properties, an EPR-like correlation is established (HBT\(\oplus\)EPR \(\rightarrow\) EPR). 
Intriguingly, this connection rule resembles an XOR logic gate, where HBT and EPR-like correlations act as logical 0 and 1, respectively, which is why we here choose the symbol \(\oplus\) as the connection symbol.
With Scheme I, Figs.~\ref{fig:7}(b1) and (b2) respectively show the momentum correlation and OAM correlation functions obtained with the correlator of Scheme I under the three cases.
For the momentum correlation function in Fig.~\ref{fig:7}(b1), when one detector selects a fixed point (non-midpoint, \(\mathbf{r}_0\)), the correlation peak of the induced HBT correlation appears at the same position \(\mathbf{r}_0\), while the correlation peak of the EPR-like correlation (2nd row) appears at the opposite position \(-\mathbf{r}_0\), i.e. momentum anti-correlation. Similarly, the OAM anti-correlation in Fig.~\ref{fig:7}(b2) (2nd row) manifests as correlation peaks appearing on the diagonal where \(\ell_a + \ell_b = 0\).
With Scheme II, Fig.~\ref{fig:7}(c1) and (c2) display the generated momentum correlation functions under the threshold conditions of \(\Delta I \leq 0.75\sigma\) and \(\Delta I > 1.5\sigma\), respectively; For the case of HBT\(\oplus\)EPR with \(\Delta I \leq 0.75\sigma\) (2nd row), both antibunching-like correlation dip (where the correlation signal is below the background) and momentum anti-correlation can be observed simultaneously.
Likewise, Figs.~\ref{fig:8}(a) and (b) respectively show the OAM correlation functions generated under two threshold conditions.
It is also possible to find a new classical correlation mode that simultaneously exhibits a correlation dip and OAM anti-correlation, and it can be flexibly manipulated by adjusting the threshold condition.

\section{Discussion}
\label{sec:3}

To conclude, we propose two realizations of the correlator for classical light, which generate optical spatial correlations between two independent uncorrelated chaotic light sources. With our schemes, we can further manipulate the bunching or anti-bunching photon correlations between two uncorrelated light sources using far fewer measurements. We also experimentally demonstrate the first GI experiment using unknown or uncorrelated light, opening a new avenue for GI experiments with chaotic light sources that are difficult to spatially resolve in time.

Of course, a prerequisite for the operation of this correlator is that the locally deployed known light and the unknown light should share as identical frequency components and polarization as possible, implying that the two beams must be capable of coherent superposition, fundamentally required in Eq.~(\ref{eq:2}). 
In principle, this classical correlator can be deployed on any propagation plane in free space, but the symmetry between different optical paths (e.g., \(A_x\), \(A_x\), \(A_x\), and \(A_x\) in Fig.~\ref{fig1}) should be maintained as much as possible. Imperfect symmetry between the optical paths and instability of the interferometer are the main sources of experimental error for this correlator. This correlator certainly remains a linear optical device and can be extended to establish correlations between two independent thermal light beams at different frequencies, based on some nonlinear optical technologies \cite{BuonoT2022,LiuLC2021,Li2025}.

In numerical simulations, the correlator is confirmed to be applicable to other spatial degrees of freedom (OAM) and to thermal light with other types of spatial correlation properties (HBT-type and EPR-like). We have also revealed the XOR-gate-like connection rule between independent thermal light sources possessing different correlation properties.
In addition, the correlator in Scheme II inherently allows for nonlocal correlation manipulation, and a total of four distinct momentum or OAM correlation modes are obtained in Figs.~\ref{fig:7} and \ref{fig:8}. This provides a novel toolkit for controlling the correlation properties in random light \cite{ChenY2022,HouW2025}.
This work mainly studied spatial correlation swapping between two independent thermal lights; in the future, we will consider extending the model to a correlation swapping network among \(N\) independent thermal light sources.

\section{Methods}\label{sec:4}

\subsection*{Experimental setup}
The experimental realization of the correlator and the interaction-free GI is shown in Fig.~\ref{fig1}.
The He-Ne laser (632.8 nm) collimated by a telescope is split into two beams by a 50/50 BS, and the two beams pass through two different RGGPs to form statistically independent pseudo-thermal light sources, A and B. The pseudo-thermal light beam A (B) is then divided into two beams, $A_x$($B_x$) and $A_y$ ($B_y$), through BS$_1$ (BS$_2$). The distances from the two thermal sources to all the detectors are the same, 0.5 m. All detectors (Daheng Imaging, MER-231-41U3M) are triggered synchronously at 8 Hz by an external signal source, and the integration time is much less than the equivalent coherence time ($\sim$5 ms) of pseudo-thermal light.
The observation method involved fixing a point (i.e., midpoint) in one detector and performing joint measurements with a region of interest in another detector to obtain a two-dimensional distribution of $g^{(2)}(\mathbf{r})$ over 750k realizations or shots.
In the interferometer setup of the correlator, the two detectors are set to bucket detection mode, i.e., they only read out the total intensity fluctuations of the region of interest.
The average size of the speckle grain of pseudo-thermal light is characterized by the width of the central peak of the two-dimensional autocorrelation distribution. Combined with the effective area of the detection field of view, the effective number \(N\) of speckles can be well estimated over 5000 realizations.

\subsection*{Numerical simulation}
In the numerical simulation of the correlator in Sections 2.2 and 2.3, the two pseudo-thermal light beams were generated by imposing pixelated random phases onto the incident Gaussian beam. Free-space Fresnel propagation was used to digitally calculate the field distributions on any observation plane. The propagation distances from the two sources to the detectors were set to be equal, both being \(0.2\,\mathrm{m}\). The wavelength was \(\lambda=632.8\,\mathrm{nm}\), the number of realizations was \(2000\mathrm{k}\), the number of transverse sampling pixels was \(46\times46\), the minimum randomization unit was \(2\times2\) pixels, and the pixel pitch was set to \(24\,\mu\mathrm{m}\). Scheme I was implemented through a third-order correlation function involving the mediator variable \(C\); Scheme II was implemented by calculating the conditional correlation functions based on the intensity difference \(\Delta I\) and its standard deviation. In the numerical simulation of the interaction-free GI experiment (Section 2.4), the number of sampling pixels was \(72\times72\), the minimum randomization unit was \(3\times3\) pixels, and the pixel pitch was \(12.5\,\mu\mathrm{m}\).

The numerical simulation results in Sections 2.5 and 2.6 are based on the setup shown in Figure~\ref{fig:6}(a). A lens-based optical Fourier transform was used to calculate the far-field distributions, with the focal length set to \(f=0.2\,\mathrm{m}\). The number of realizations was \(2000\mathrm{k}\), the number of sampling pixels was \(46\times46\), the minimum randomization unit was \(2\times2\) pixels, and the pixel pitch was \(24\,\mu\mathrm{m}\). The helical phases \((\ell_a,\ell_b)\) carrying the topological charge from \(-5\) to \(+5\) were loaded in the SLMs. These SLMs were located in the far field of RGGPs. 
For each pair of helical phases loaded onto the two SLMs, the two detectors were used to calculate the OAM correlation function \(g^{(2)}(\ell_a,\ell_b)\) (11\(\times\)11) between the two projected optical fields.
The two detectors were placed at the far-field plane of the SLMs, and only read out the intensity fluctuations at the midpoint for correlation calculation.
In contrast, the other two detectors in the interferometer setup of the correlator still perform bucket detection mode on the region of interest. 
The various simulation parameters in Section 2.6 are the same as those in Section 2.5; the only thing that needs to be changed is the cross-correlation property of the pairwise input random optical fields. HBT-type and EPR-like pseudo-thermal light beams are simulated, respectively, by a pair of random optical fields with identical random phases and a pair of random optical fields with complementary random phases at the source plane. 

\backmatter

\bmhead{Supplementary information}
Please refer to the Supplementary Material for the complete visualization data and detailed theoretical derivation.

\begin{itemize}
\item Funding: We wish to acknowledge the support of the National Natural Science Foundation of China (12274037, 12504411, 11735005, 11654003), Basic and Applied Basic Research Foundation of Guangdong Province (2026A1515011704), the Science and Technology Development Fund from Macau SAR (FDCT) (0105/2023/RIA2), the research outputs of this Open Research Project Programme are funded by the Science and Technology Development Fund of the Macao S.A.R. (File No.: 0002/2024/TFP), and the Fundamental Research Funds for the Central Universities.
\item Conflict of interest: The authors declare no conflict of interest. 
\item Data availability: Experimental and numerical data are available upon reasonable request.
\item Author contribution: Z.Y. and K.W. conceptualized the study and wrote the manuscript. Z.Y. and J.X. designed the experiments. W.H. constructed the experiment and made the figures with help from J.Z., H.C.L., and H.B.W. K.W. performed the theory and analytics. All authors analyzed the data and edited the manuscript. J.X. supervised the project.
\end{itemize}

\end{document}